\documentclass[iop]{emulateapj}
\usepackage{graphicx}
\usepackage{subfigure}
\usepackage{hyperref}
\usepackage{epsfig}
\usepackage{natbib}
\usepackage{multirow}
\usepackage{breakurl}

\newcommand{\Text}[1]{\text{\textnormal{#1}}} 

\begin{document}

\title{Quantifying Mass Segregation and New Core Radii for 54 Milky Way Globular Clusters}

\author{ Ryan ~Goldsbury\altaffilmark{1}, Jeremy ~Heyl\altaffilmark{1}, Harvey ~Richer\altaffilmark{1}}

\altaffiltext{1}{Department of Physics \& Astronomy, University of British Columbia, Vancouver, BC, Canada V6T 1Z1; rgoldsb@phas.ubc.ca, heyl@phas.ubc.ca, richer@astro.ubc.ca}


\begin{abstract}

We present core radii for 54 Milky Way globular clusters determined by fitting King-Michie models to cumulative projected star count distributions.  We find that fitting star counts rather than surface brightness profiles produces results that differ significantly due to the presence of mass segregation.  The sample in each cluster is further broken down into various mass groups, each of which is fit independently, allowing us to determine how the concentration of each cluster varies with mass.  The majority of the clusters in our sample show general agreement with the standard picture that more massive stars will be more centrally concentrated.  We find that core radius vs. stellar mass can be fit with a two parameter power-law.  The slope of this power-law is a value that describes the amount of mass segregation present in the cluster, and is measured independently of our distance from the cluster.  This value correlates strongly with the core relaxation time and physical size of each cluster.  Supplementary figures are also included showing the best fits and likelihood contours of fit parameters for all 54 clusters.

\end{abstract} 

\section{Introduction}

Comprehensive catalogues of Milky Way globular cluster parameters have been compiled by many different groups over the last two decades.  The first large scale effort was put together by \cite{trager-1995-aj}, in which they presented surface brightness profiles for 125 Galactic globular clusters, and included parameters determined from model fitting for 63 of them.  This work was expanded upon by \cite{mclaughlin-2005-apjs} who used data from the previous paper as well as new data, and fit multiple classes of models to the surface brightness profiles.  The results from both of these papers, as well as many others, have been compiled in \cite{harris-1996-aj} (2010 edition).  Our approach differs from these previous studies in that we do not assume a single mass-to-light ratio for the cluster.  We instead work directly from star counts and consider bins 0.8 magnitudes wide to break each cluster down further, fitting the distribution of stars in each bin independently.  Using stellar evolution models from \cite{dotter-2008-apjs} we can assign masses to each bin and analyze how the stellar concentration changes with mass.  This approach avoids assuming a constant mass-to-light ratio for a cluster.  Given the presence of any amount of mass segregation the assumption of constant mass-to-light ratio will not be true since mass-to-light ratio is a function of mass, and the average mass of stars changes as a function of distance from the cluster core.  Because of this, this surface brightness profile does not necessarily reflect the underlying stellar density in a cluster.

\section{Data}

All of the data used in this study are from the ACS Survey of Galactic Globular Clusters \citep{sarajedini-2007-aj}.  A thorough discussion of the reduction can be found in \cite{anderson-2008-aj}.  Reduced catalogues of both real and artificial stars can be found at: \url{http://www.astro.ufl.edu/~ata/public_hstgc/databases.html} .

\section{Methods}

There is some ambiguity in the literature regarding a few parameters commonly used to characterize the concentration of star clusters.  In an attempt to avoid this, we will use the following symbols and definitions for the remainder of the paper.

\begin{itemize}
  \item core radius: $R_c$
  \item King radius: $r_0$
  \item tidal radius: $r_t$
  \item concentration: $c$
\end{itemize}

The latter three quanitites are defined as follows:

\begin{equation}
\Sigma(R_c)=\frac{1}{2}\Sigma(0)
\end{equation}

\begin{equation}
r_{0}=\sqrt{\frac{9\sigma ^2}{4\pi G \rho _0}}
\end{equation}

\begin{equation}
c=\log_{10}(\frac{r_t}{r_0})
\end{equation}

The quantity $\Sigma$ here refers to the projected density distribution of the cluster.  The core radius ($R_c$) is the projected radial distance from the center of the cluster at which the projected density of the cluster drops to half of the central value.  The quantity $\sigma$ is the velocity-dispersion parameter, and $\rho_0$ is the central density of the King model.  The King radius ($r_0$) can be calculated from the previous two quanities and describes the scale of the model similar to the core radius.  The concentration ($c$) is defined following the convention in \cite{bin-trem}  (hereafter BT87).  This disagrees with the definition used by \cite{harris-1996-aj}, in which $c=\log_{10}(r_t/R_c)$.  For a given $r_0$ and $r_t$ a King model has only one possible value of $R_c$, but it is not equivalent to $r_0$.  It is also important to note here that we use the capital $R$ to denote a projected radius, while the lower-case $r$ refers to a three-dimensional radius.  Equation 1 is intentionally ambiguous as it does not indicate whether the surface density is in units of luminosity, mass, or number of stars.  This will be expanded upon in Section \ref{lumbias}.

\subsection{Solving for Projected Density Distributions of King-Michie Models}

Our method involves generating a number of King-Michie \citep{king-1966-aj,michie-1963-mnras} models over varying $\Psi(0)/\sigma^2$, which are defined in Equations 4 and 5.  $\Psi(0)/\sigma^2$ is referred to as the central dimensionless potential and is sometimes written $\Psi_0$.  We follow the prescription in Section 4.4 of BT87, ending with a series of models giving normalized surface density ($\Sigma/(\rho_0 r_0)$) as a function of radius ($r/r_0$).

To calculate these models, we begin with Equation 4-132 from BT87, reproduced in Equation 4.  We have two boundary conditions.  The first is always the same: $d\Psi/dr=0$ when $r=0$.  The second is the value of $\Psi(0)$, which determines the concentration of the resulting density model for a given $\sigma$.  Since we are only concerned with $\Psi(0)/\sigma^2$, which will control the shape of the density distribution, and not $\Psi(0)$ or $\sigma$ independently, which determine the absolute scale of the system, we only consider $\Psi(0)/\sigma^2$ as a single combined parameter.

We use a fourth order Runge-Kutta method (RK4) to solve this equation numerically.  The step size varies from $0.01 r_0$ for the most concentrated to $10^{-4}r_0$ for the most extended models in our grid.  The truncation error is less than $0.1\%$ for all models calculated.  A solution to this equation for a given set of boundary conditions gives $\Psi(r)$.  This can then be fed into Equation 4-131 from BT87 (shown as Equation 5 in this paper) to go from $\rho(\Psi)$ to $\rho(r)$.

break

\begin{widetext}

\begin{equation}
\frac{d}{dr}\left(r^2 \frac{d\Psi }{dr}\right)=-4 \pi G \rho_1 r^2 \left[e^{\Psi / \sigma^2}\Text{erf} \left(\frac{\sqrt{\Psi}}{\sigma} \right)-\sqrt{\frac{4\Psi}{\pi \sigma^2}} \left(1+\frac{2\Psi}{3\sigma^2} \right)\right]
\end{equation}

\begin{equation}
\rho(\Psi)=\rho_1 \left [ e^{\Psi / \sigma^2} \Text{erf} \left(\frac{\sqrt{\Psi}}{\sigma} \right) - \sqrt{\frac{4 \Psi}{\pi \sigma^2}} \left(1 + \frac{2 \Psi}{3 \sigma^2} \right ) \right ]
\end{equation}

\end{widetext}

Finally, to transform from a three dimensional density distribution to a projected density distribution, one must perform an Abel Transformation:

\begin{equation}
\Sigma (R) = \int_{R}^{\infty} \frac{\rho(r) r dr}{\sqrt{r^2-R^2}}\ \ .
\end{equation}

\begin{figure}[h!tbp]
\centering
\epsfig{file=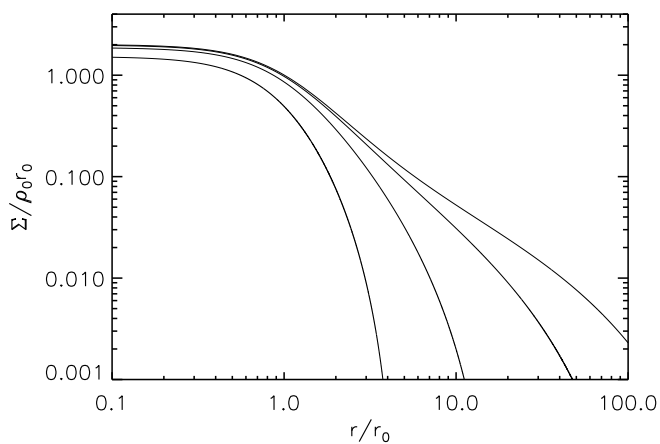,width=\linewidth}
\caption{We have recalculated Figure 4-9b from BT87, showing the projected density distribution for various boundary conditions.  These distributions correspond to (from left to right) $\Psi(0) / \sigma^2 = 12,9,6,3$ or $c=2.74,2.12,1.26,0.67$ since there is a one-to-one relation between $\Psi(0) / \sigma^2$ and $c$.}
\label{projden}
\end{figure}

A range of the resulting projected density distributions is shown in Figure \ref{projden}.  The brief summary in this section is not intended to be comprehensive, only to allow the reader to replicate our procedure for calculating these models.  A more thorough description, including motivation for the distribution function from which these models are derived, can be found in \cite{king-1966-aj} or BT87.

\subsection{Luminosity vs. Star Counts}
\label{lumbias}

\begin{figure*}[htbp]
\centering
\includegraphics[scale=1]{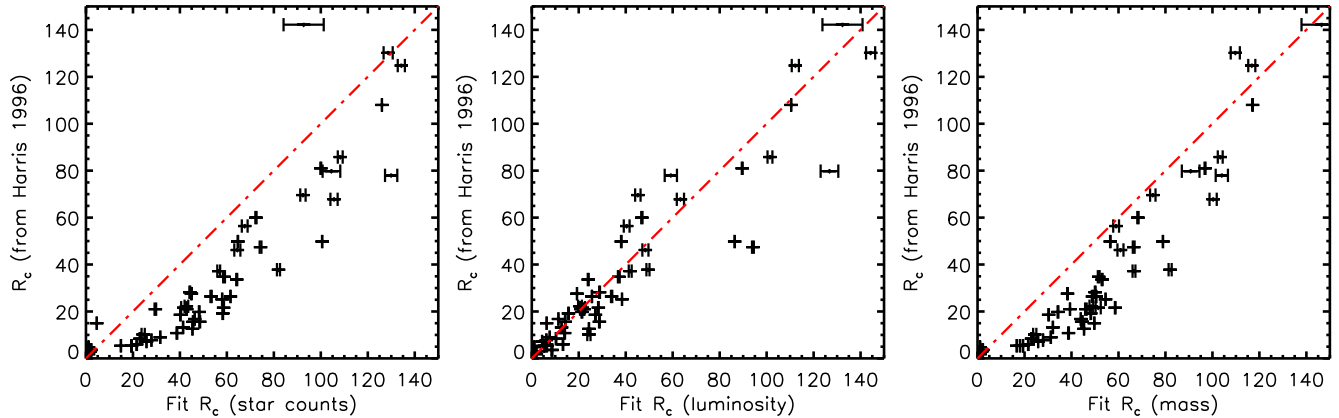}
\caption{The left hand panel shows core radius values taken from \cite{harris-1996-aj} plotted against values determined using our approach of fitting the cumulative star count distribution.  The center panel shows values determined by fitting the cumulative distribution in which objects are weighted by their luminosities.  The right panel shows fits to the mass weighted cumulative distributions.  The red line highlights a non-biased relation in each case.  The horizontal error bars represent 1$\sigma$ and the Harris values have no error plotted.}
\vspace{20pt}
\label{rvharr}
\end{figure*}

In the interest of presenting a straightforward comparison to previously determined values of $R_c$ we first fit the distribution of all stars in a given cluster field together.  While our grid is parametrized by $r_0$, it is easy to determine the value of $R_c$ for a given model by finding the radius at which the surface density drops to half.  This value is more commonly referenced in the literature.  Our fitting method will be discussed in Section \ref{fitmodels}.  A comparison between the values currently in the literature and our resulting best fit values and their uncertainties is shown in the left hand panel of Figure \ref{rvharr}.  Our uncertainties are estimated with an iterative fitting procedure that considers uncertainty in the location of the cluster center as well as uncertainty due to sample size through bootstrapping.  We find that by fitting to the cumulative star count distribution and assuming a conservative $20\%$ uncertainty in Harris' values the mean difference between our fit values and Harris' values is inconsistent with zero at a bit more than $4\sigma$, where the error used is the standard error.  We tend to measure statistically significant larger core radii.  This is not surprising, given that $R_c$ values reported in \cite{harris-1996-aj} are determined from fitting the luminosity profiles of clusters.    

If we repeat our fitting procedure, but now weight each star's contribution to the cumulative distribution by its luminosity, we then produce the result shown in the center panel of Figure \ref{rvharr}.  Here the bias described above is removed and the mean difference between our values and those of Harris is within $1.2\sigma$ of zero.

The explanation for this discrepancy is as follows.  The mass-to-light ratio is a function of radius in most globular clusters due to mass segregation; more massive (and therefore more luminous) stars will be more centrally concentrated than less massive (less luminous) stars.  This implies that the mass-to-light ratio decreases toward the center of the star cluster and so converting between a distribution measured from surface brightness to the underlying distribution of stars is not a constant factor.  In fact, how this mass-to-light ratio changes with radius also differs between clusters.  So, even if we consider mass-to-light ratio as a function of $R$, there is no universal correction to be applied.

If we wish to determine the mass distribution of the cluster, then fitting the star count distribution still has an inherent bias for the same reason discussed above.  We make a rough approximation of the core radius of the true underlying mass distribution by fitting cumulative distributions in which stars are weighted by mass.  The masses are assigned by fitting isochrone models from \cite{dotter-2008-apjs} to the cluster CMDs.  The results are shown in the right panel of Figure \ref{rvharr}.  These values fall somewhere in between the previous two, but the key point is that they are inconsistent with those determined from fitting the surface brightness profile.  This indicates that the surface brightness profile is not an appropriate proxy for the mass surface density of a cluster.  For the remainder of the paper we will attempt to quantify these effects by analyzing how subgroups of varying mass are distributed in each cluster.

\subsection{Selecting Groups Along the Main-Sequence}

\begin{figure}[h!tbp]
\centering
\epsfig{file=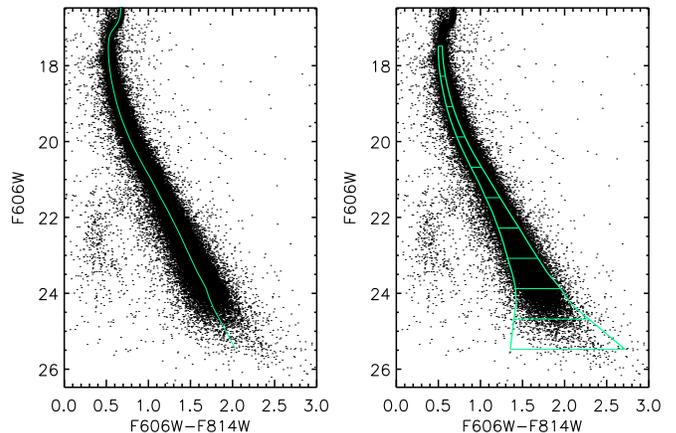,width=\linewidth}
\caption{The algorithmically determined fiducial sequence and the regions for the ten selected bins are shown for the cluster 47 Tucanae (NGC 104).}
\label{fidplot}
\end{figure}

For each cluster, we create ten independent groups along the main sequence.  We begin by fitting a fiducial sequence to each cluster color-magnitude diagram (CMD) in $F606W$ vs. ($F606W-F814W$).  The fiducial is created using an iterative algorithm.  A manual starting point is selected at the base of the CMD.  The algorithm then chooses a direction to move by using the AMOEBA algorithm from \cite{numrec} to minimize the median of the absolute separation of the nearby points perpendicular to each proposed step.  There is also an additional cost added based on the angle between sequential steps to prevent the sequence from bending back on itself.  The result is a fiducial that traces out the center of the sequence.  Once this fiducial is established for each cluster, the turn-off is then defined as the magnitude that produces the most negative derivative ($dcol/dmag$) of the fiducial.  Ten magnitude bins are created starting from 0.5 magnitudes below this turn-off and covering eight magnitudes in total.  The widths of the bins are determined by calculating the standard deviation of the CMD in the color direction from the fiducial.  Only objects that fall within $\pm3\sigma$ are included.  While this approach does require an initial point to be selected manually on the CMD, it was found that various starting points would converge to the same sequence within approximately 5 iterations.  These initial burn-in points were removed and the fiducial was linearly interpolated backward from the later points.  This effectively removes all human input, leading to a bin selection method that is entirely automated and consistent across all of the clusters used in this study.  Figure \ref{fidplot} illustrates the location of these bins on a CMD.

\subsection{Fitting Models to Subgroups}
\label{fitmodels}

For each cluster, we fit our grid of King-Michie models to the cumulative radial distribution of each of our 10 sub-groups.  We fit only out to projected radii of 100 arc-seconds, due to the size limitations of a single ACS field.  Out to 100 arc-seconds from the center ($R_{max}$ in Equation \ref{cdist}), each annulus is fully within the field.  Beyond this, more and more of each successive annulus will fall off of the field.  It is possible to take this effect into account when fitting the models, but including these regions does not significantly improve our ability to constrain the best fitting model parameters, and so they are ignored in the interest of simplicity.

To generate a model to fit to our data we need to specify two of the three parameters: $r_0$, $r_t$, and $c$, since the third can always be determined from the other two.  The tidal radius of the cluster in all cases will be many times the maximum radius that can be observed with our field, and we expect this to be a constant for all subgroups within the cluster.  Therefore, instead of fitting for two parameters, we use the values of $c$ and $R_c$ from \cite{harris-1996-aj} to calculate $r_t$ for each cluster and hold this value fixed.  

Before fitting the models to the data, we also need to take into account the completeness of the stars in each bin.  To do this, we use the artificial star tests mentioned in Section 2 to calculate the completeness as a function of radius for each sub-group in each cluster.  The incompleteness ($I$) is the probability that a star of a given magnitude would be found at a given position in the field.  These corrections are then applied to the model density distributions before fitting to the data.

Given the projected density distribution $\Sigma(R)$ and data incompleteness $I(R)$, we calculate the cumulative distribution as:

\begin{equation}
C(R)=\frac{\int_{0}^{R}\Sigma(R')I(R')R' dR'}{\int_{0}^{R_{max}}\Sigma(R')I(R')R' dR'} .
\label{cdist}
\end{equation}

The angular component of this integral is not shown since our models are spherically symmetric and therefore azimuthally symmetric in projection, and the result simplifies to the equation shown above.  Here $I$ is only dependant on $R$ since we are considering a group of objects that fall within a narrow magnitude range.

In determining the best fit to the cumulative distribution of each bin, there are two major sources of uncertainty.  The first is the location of the center of the cluster.  We use the center coordinates and uncertainties from \cite{goldsbury-2010-aj}, which were determined using this same data set.  The second source of uncertainty is the sample size in each bin.  Both of these are taken into account by using a bootstrap fitting method as follows.  In each iteration, a sample of the same size as the full bin sample is chosen with replacement.  The location of the center to use is drawn from a two dimensional Gaussian distribution centered at the value from \cite{goldsbury-2010-aj} with the appropriate $\sigma$.  The cumulative radial distribution is calculated and the best fitting model is determined by minimizing the maximum separation between the real distribution and the model, which is the Kolmogorov-Smirnov statistic $D$.  In each iteration, we record a single value for the best fitting $r_0$.  After many iterations we can then generate a histogram of fit $r_0$ values.  This distribution is roughly Gaussian and is broadened by both sources of error discussed above.  We take the mean and standard deviation of these values as our best fitting $r_0$ and $1\sigma$ uncertainty for each sub-group in each cluster.   The dominant source of error in most cases is the $\sqrt{N}$ random uncertainty.  However, in very well populated bins, such as the low mass groups in a very massive cluster like NGC 5139, the systematic uncertainty from the location of the cluster center contributes as much to the total error as the Poisson counting uncertainty.

\subsection{Assigning Masses to Bins and Fitting Power Laws}
\label{fitr0}

After iterating the fitting procedure described in Section \ref{fitmodels} for all sub-groups in all clusters, we can show how the concentration of stars along the main-sequence changes with magnitude for each cluster.  However, this is not particularly useful for making comparisons among clusters, since the apparent magnitude of an object is a result of a number of factors: distance, extinction, mass, metallicity, and cluster age.  In terms of dynamics, we would like to look at how the concentration changes with object mass.  So, to assign a mass to each magnitude bin, we need to consider the effects of the other four parameters.  

To do this, we again use an iterative approach.  In each iteration we draw values of the four parameters listed above from Gaussian distributions.  The parameters of these distributions are drawn from a number of sources.  All of the distances, metallicities, and extinctions are taken from \cite{harris-1996-aj}.  The vast majority of these have no reported uncertainties in the original references, and so we assume conservative estimates: for distance we assume a standard deviation of 0.2 magnitudes, for extinction and metallicity we assume $20\%$ uncertainties.  The ages and uncertainties are taken from \cite{dotter-2010-apj} with the exception of five clusters that were not listed in that paper.  These are: NGC 1851 and NGC 2808 \citep{koleva-2008-mnras}, NGC 5139 \citep{forbes-2010-mnras}, NGC 6388 \citep{catelan-2006-apjl}, and NGC 6715 \citep{geisler-2007-pasp}.

  These parameters are then used to define a stellar isochrone model from \cite{dotter-2008-apjs}.  For each magnitude bin, we assign masses to all of the objects using the model and then take the mean mass of all objects in each bin.  So, for each iteration, we record one mass value for each bin.  We then use these mass values along with our best fitting $r_0$ value and uncertainties to fit a two parameter power-law model to the relation between King radius and mass using a maximum likelihood approach.  The isochrone model grid does not agree well with the structure of most CMDs at the very low mass end, so we ignore all bins for which the mean mass is less than 0.2 $M_\odot$.  We also ignore points for which the best fitting $r_0$ falls outside of our field.  In these cases, $r_0$ is not well constrained by our data.  For each iteration, we save the likelihood results over the two power-law parameters shown in Equation \ref{pwrlaw_eq}. This power-law model is not physically but rather empirically motivated, as it fits a wide range of clusters well with a small number of parameters:

\begin{equation}
r_{0}=A\left(\frac{M}{M_\odot} \right)^B .
\label{pwrlaw_eq}
\end{equation}

\begin{figure}[htbp]
\centering
\epsfig{file=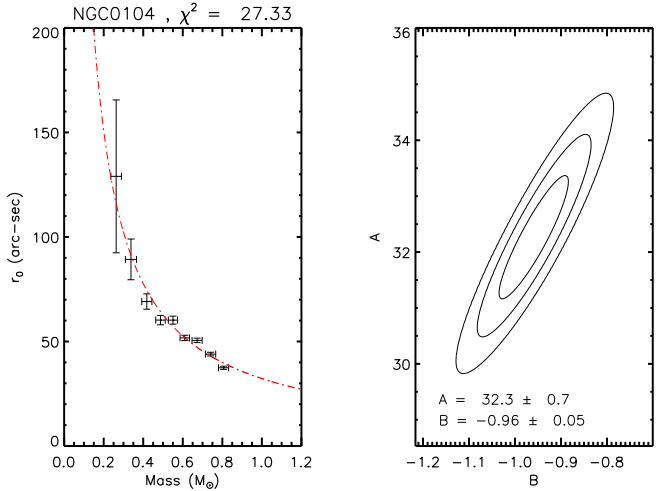,width=\linewidth}
\caption{The best fitting power-law model and likelihood of fit parameters for 47 Tuc.  Similar figures can be found in the supplement showing power-laws fit to both $r_0$ and $R_c$ for all 54 clusters in our sample.}
\label{pwrfit}
\end{figure}

After many iterations as described above, we can average this stack of two dimensional likelihood surfaces.  This amounts to a Monte Carlo integration over the likelihood of the four other parameters (distance, extinction, metallicity, and cluster age) with prior distributions on each parameter.  This leaves the likelihood of only the two power-law parameters, but with the uncertainties in the other four propagated through.  An example fit for 47 Tuc is shown in Figure \ref{pwrfit}.  

\section{Results}

The parameter $A$ is a normalization that corresponds to the King radius of cluster members at 1 solar mass.  This value is given in projected coordinates which are related to the true three dimensional King radius by the distance to the cluster.  The parameter $B$ describes how much the concentration of stars changes with their mass, and is independent of the total scale of the cluster.  Although we choose to fit our models for $r_0$, we could just as easily parametrize them by $R_c$ and fit power-laws to this parameter in the same way as Equation \ref{pwrfit}.  For all 54 of the clusters included in this analysis, these two power-law parameters as well as their uncertainties and the $\chi^2$ values for the best fit (the value given is the total $\chi^2$, not per degree of freedom) are included in Table 2 for both $r_0$ and $R_c$ as a function of mass.  For any of these clusters, this empirical relation can then be used to estimate the distribution of stars of a given mass.  The values of $A$ in this table are in units of arc-seconds, and the parameters correspond to a power-law function in the form of Equation \ref{pwrlaw_eq}, but in the right side the parameters are for fits to $R_c(M)$ rather than $r_0(M)$.  Although $r_0$ is determined from the three dimensional density distribution, our values come from fitting the projected density distribution and so we fit in units of angle rather than distance.  Since we are fitting in projected space, we leave our results in these projected units.  These results can be converted to physical units by using the distance given in \cite{harris-1996-aj} or another source.

\begin{figure*}[!htbp]
\centering
\epsfig{file=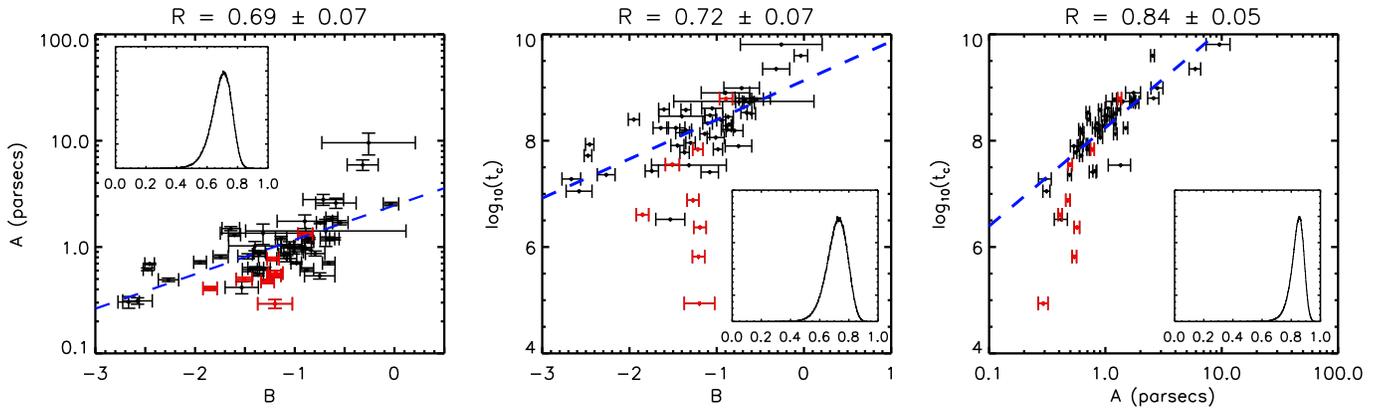,width=\linewidth}
\caption{Correlations between our fit parameters and the core relaxation time of the cluster.  Here $A$ has been scaled by the distance to each cluster so that it is in units of parsecs.  Red points indicate core collapsed clusters not used in calculating the correlation statistic.  The insets show the histogram of measured Pearson correlation coefficients after bootstrapping and taking into account uncertainties in the fit parameters.}
\label{corr_plot}
\end{figure*}

Parameter $A$ has units of angle and our measurement of it depends on our distance from the cluster.  Parameter $B$, on the other hand,  is unitless and would be measured the same regardless of our distance from the cluster.  This parameter is a quantitative descriptor of the mass segregation present in a cluster.  We can use the apparent distance modulus and extinction from \cite{harris-1996-aj} to convert our fit values for $A$ in arc-seconds (or pixels) to parsecs, which gives us a parameter to describe the intrinsic size of the cluster.  After doing this, we find that $A$ and $B$ correlate strongly with each other such that larger clusters tend to have less mass segregation than small clusters.  Both of these quantities also correlate with the core relaxation time from \cite{harris-1996-aj}.  Figure \ref{corr_plot} shows these correlations.

The distribution inset in each diagram indicates the Pearson correlation coefficient ($R$) distribution determined through bootstrapping.  A principal component analysis of this three-dimensional space indicates only a single significant component that links all three dimensions.  This can be understood as follows: large clusters have longer crossing times and therefore longer relaxation times.  The amount of mass segregation present in the cluster should approach some limit near equipartition as many relaxation times pass.  Therefore, these larger and slower to relax clusters will exhibit less mass segregation than those that are small and quick to relax.  

Lines of the following forms are fit to each correlation.  

\begin{eqnarray}
\log_{10}(A)=b_1+S_1*B \\
\log_{10}(t_c)=b_2+S_2*B \\
\log_{10}(t_c)=b_3+S_3*\log_{10}(A)
\label{linfiteq}
\end{eqnarray}

These fits are shown as blue dashed lines in Figure \ref{corr_plot}, and the values of the best fit parameters as well as their variances and covariances are listed in Table 1.  As with the calculation of the correlation coefficients, core collapsed clusters are excluded from these fits.  Classification of core collapsed clusters is taken from \cite{harris-1996-aj}.  These are derived from luminosity profiles and so this classification potentially suffers from the biases introduced by mass segregation that are discussed above.  Of the eight clusters that are classified as core collapsed, we find that the projected density distribution measured from star counts is well modeled by just a powerlaw for six of them.  Two of the clusters (NGC 6541 and NGC 6723) show clear evidence of leveling off toward the center and do not appear to have collapsed cores when the density is measured through star counts rather than luminosity.  Similar discrepancies have been found by \cite{miocchi-2013-arx}, where clusters are found to show a central cusp in their surface brightness profile but no such feature in their stellar density profile.  The authors attribute this to the presence of a few anomalous bright stars, but we believe this discrepancy could also be caused by the systematic effects of mass segregation.  Despite the fact that clusters which appear to be core collapsed in surface brightness may not actually be so, we have still excluded all of these clusters from the correlation analysis shown in Figure \ref{corr_plot} since this classification still carries a potential bias through the calculation of the relaxation time.

\begin{center}
\begin{table}[!htbp]
\caption{Fit Correlations}

\centering
\begin{tabular}{|c || c | c || c | c | c |}

\hline
Relation & b & S & Var(b) & Var(S) & COV(b,S) \\
\hline

$\log_{10}(A)$ vs. $B$              &  0.35 &  0.29 &  0.007 &  0.003 &  0.005\\
$\log_{10}(t_c)$ vs. $B$            &  9.06 &  0.68 &  0.019 &  0.012 &  0.013\\
$\log_{10}(t_c)$ vs. $\log_{10}(A)$ &  8.25 &  1.84 &  0.003 &  0.043 & -0.001\\

\hline
\end{tabular}
\label{line_table}
\end{table}
\end{center}

\section{Conclusions}

We have shown that measuring the core radius of a cluster through the surface brightness introduces a bias that makes clusters appear more concentrated than the actual distribution of their main sequence stars or total mass on the main sequence.  This bias is explained by the presence of mass segregation within clusters.  Mass segregation creates a mass-to-light ratio that changes with radius, and so there is no constant factor that can be used to convert between the density distribution of a cluster as measured through the surface brightness profile and the true underlying stellar density.  Cluster model parameters measured in this way will be biased to the most massive stars in the cluster, and will not accurately describe the distribution of lower mass stars.  Such a discrepancy was also recently 

We have attemped to quantify the amount of mass segregation in 54 clusters using public catalogs \citep{sarajedini-2007-aj}.  This was done by fitting King-Michie models to a series of narrow mass bins in each cluster, allowing us to measure how the King radius varies with mass.  We found that a simple two parameter power-law fit these data well and we have presented the parameters of these fits for all of the clusters in Table 1.  We also present the best fitting power-law and the likelihood contours of fit parameters for each cluster in the supplementary material.  We have shown that a single concentration parameter does not accurately reflect the distribution of most stars within a cluster, and the models we have presented could be used to estimate the distribution of stars as a function of mass within the clusters.

We find that the exponent of our power-law model correlates strongly with both the core relaxation time and the absolute size of the cluster.  This suggests that these models could be used as a proxy for the underyling dynamical state of the cluster.  However, the scatter in these correlations is larger than can be attributed to the errors alone.  In the case of the second and third correlations, this is partly attributable to relaxation time values having no associated uncertainty.  However, the scatter still appears to be larger than expected in the first correlation, for which uncertainties have been properly considered.  This potentially indicates the effects of other factors, not considered in this work, that can influence the dynamical evolution of clusters.

\begin{center}
\begin{table*}
\caption{Fit Powerlaw Parameters}

\centering
\begin{tabular}{|c || c | c | c || c | c | c |}

\hline
 & \multicolumn{3}{| c ||}{$r_0$} & \multicolumn{3}{| c |}{$R_c$} \\
\hline
Cluster & A & B & $\chi^2$ & A & B & $\chi^2$ \\
\hline

NGC0104 &  32.3$\pm$ 0.7 & -0.96$\pm$0.05 &  27.3 &  24.4$\pm$ 0.6 & -0.95$\pm$0.05 &  27.1\\
NGC0288 &  64.9$\pm$ 7.6 & -0.67$\pm$0.19 &   2.5 &  48.5$\pm$ 5.7 & -0.58$\pm$0.19 &   2.0\\
NGC1261 &  16.9$\pm$ 0.4 & -1.55$\pm$0.03 &  13.1 &  14.2$\pm$ 0.5 & -1.15$\pm$0.03 &  32.8\\
NGC1851 &  15.6$\pm$ 0.5 & -1.74$\pm$0.07 &  50.9 &  12.0$\pm$ 0.4 & -1.61$\pm$0.07 &  59.1\\
NGC2298 &  11.7$\pm$ 0.5 & -1.54$\pm$0.08 &   9.4 &   8.9$\pm$ 0.4 & -1.48$\pm$0.08 &   9.9\\
NGC2808 &  17.0$\pm$ 0.9 & -1.48$\pm$0.11 &  37.7 &  11.9$\pm$ 0.3 & -1.60$\pm$0.02 &  38.8\\
NGC3201 &  44.8$\pm$ 2.7 & -0.96$\pm$0.11 &  12.6 &  33.8$\pm$ 2.0 & -0.93$\pm$0.11 &  12.3\\
NGC4147 &   8.3$\pm$ 0.6 & -1.11$\pm$0.10 &  35.3 &   6.3$\pm$ 0.4 & -1.09$\pm$0.07 &  34.3\\
NGC4590 &  23.0$\pm$ 0.7 & -0.89$\pm$0.05 &   6.8 &  17.3$\pm$ 0.6 & -0.87$\pm$0.05 &   6.8\\
NGC4833 &  38.7$\pm$ 1.3 & -0.61$\pm$0.05 &   7.7 &  29.0$\pm$ 1.0 & -0.59$\pm$0.05 &   7.5\\
NGC5024 &  19.6$\pm$ 0.4 & -0.74$\pm$0.03 &  11.4 &  14.8$\pm$ 0.3 & -0.73$\pm$0.03 &  11.2\\
NGC5053 & 127.9$\pm$40.6 & -0.24$\pm$0.62 &   1.9 &  81.5$\pm$20.0 & -0.39$\pm$0.37 &   1.9\\
NGC5139 & 102.2$\pm$ 4.9 & -0.06$\pm$0.06 &   3.0 &  76.7$\pm$ 3.6 & -0.06$\pm$0.06 &   2.9\\
NGC5272 &  24.4$\pm$ 0.5 & -0.88$\pm$0.04 &  20.3 &  18.4$\pm$ 0.4 & -0.87$\pm$0.04 &  20.2\\
NGC5286 &  13.0$\pm$ 0.4 & -1.96$\pm$0.06 &  19.2 &  10.1$\pm$ 0.3 & -1.83$\pm$0.06 &  23.4\\
NGC5466 &  76.6$\pm$ 8.8 & -0.31$\pm$0.15 &   3.8 &  56.4$\pm$ 6.5 & -0.29$\pm$0.15 &   3.2\\
NGC5904 &  26.7$\pm$ 0.7 & -0.91$\pm$0.05 &  26.5 &  20.1$\pm$ 0.5 & -0.90$\pm$0.05 &  26.2\\
NGC5927 &  27.4$\pm$ 0.6 & -1.00$\pm$0.05 &   7.7 &  20.6$\pm$ 0.5 & -0.98$\pm$0.05 &   7.1\\
NGC5986 &  18.2$\pm$ 0.5 & -1.24$\pm$0.05 &  18.2 &  13.9$\pm$ 0.4 & -1.17$\pm$0.05 &  19.5\\
NGC6093 &  12.6$\pm$ 0.4 & -1.17$\pm$0.05 &  61.8 &   9.5$\pm$ 0.3 & -1.13$\pm$0.05 &  61.0\\
NGC6121 &  50.8$\pm$ 3.2 & -0.61$\pm$0.12 &   5.9 &  38.3$\pm$ 2.4 & -0.60$\pm$0.12 &   5.9\\
NGC6144 &  40.4$\pm$ 1.6 & -0.49$\pm$0.06 &  10.6 &  30.5$\pm$ 1.2 & -0.49$\pm$0.06 &  10.5\\
NGC6171 &  29.3$\pm$ 1.3 & -0.88$\pm$0.08 &  13.8 &  22.1$\pm$ 1.0 & -0.86$\pm$0.08 &  13.6\\
NGC6205 &  35.3$\pm$ 0.8 & -0.65$\pm$0.05 &  13.6 &  26.5$\pm$ 0.6 & -0.64$\pm$0.05 &  13.3\\
NGC6218 &  37.5$\pm$ 1.8 & -0.76$\pm$0.08 &  10.7 &  28.2$\pm$ 1.4 & -0.74$\pm$0.08 &  10.4\\
NGC6254 &  30.7$\pm$ 1.0 & -0.64$\pm$0.04 &  12.4 &  23.1$\pm$ 0.7 & -0.62$\pm$0.04 &  12.2\\
NGC6304 &  16.4$\pm$ 0.7 & -2.18$\pm$0.10 &   6.6 &  13.7$\pm$ 0.3 & -1.85$\pm$0.03 &  16.4\\
NGC6341 &  16.9$\pm$ 0.5 & -1.12$\pm$0.04 &  26.4 &  12.8$\pm$ 0.3 & -1.10$\pm$0.04 &  26.7\\
NGC6352 &  36.6$\pm$ 3.8 & -1.36$\pm$0.26 &   3.6 &  28.6$\pm$ 1.7 & -1.16$\pm$0.07 &   3.7\\
NGC6362 &  70.3$\pm$ 7.0 & -0.68$\pm$0.19 &   8.9 &  52.1$\pm$ 5.1 & -0.60$\pm$0.19 &   6.5\\
NGC6366 &  96.0$\pm$23.1 & -0.57$\pm$0.41 &   2.1 & 109.3$\pm$11.4 & -0.30$\pm$0.16 &  16.8\\
NGC6388 &  13.7$\pm$ 0.4 & -2.39$\pm$0.04 & 183.3 &  12.3$\pm$ 0.2 & -1.84$\pm$0.03 &  79.0\\
NGC6397 &  28.7$\pm$ 2.2 & -1.01$\pm$0.13 &   6.5 &  21.7$\pm$ 1.7 & -0.98$\pm$0.13 &   6.2\\
NGC6441 &  12.9$\pm$ 0.2 & -2.18$\pm$0.04 & 544.3 &  10.6$\pm$ 0.2 & -1.82$\pm$0.04 & 261.1\\
NGC6535 &   9.7$\pm$ 1.2 & -2.47$\pm$0.10 &   8.6 &   9.5$\pm$ 1.1 & -1.89$\pm$0.10 &   5.1\\
NGC6541 &  14.3$\pm$ 0.6 & -1.44$\pm$0.07 &  47.7 &  10.9$\pm$ 0.4 & -1.41$\pm$0.07 &  47.9\\
NGC6584 &  18.2$\pm$ 0.7 & -1.26$\pm$0.06 &  11.0 &  13.8$\pm$ 0.5 & -1.20$\pm$0.06 &  11.4\\
NGC6624 &  10.6$\pm$ 0.4 & -1.82$\pm$0.07 &  13.0 &   8.0$\pm$ 0.3 & -1.80$\pm$0.07 &  12.3\\
NGC6637 &  14.9$\pm$ 0.5 & -1.40$\pm$0.08 &   6.6 &  11.3$\pm$ 0.4 & -1.35$\pm$0.08 &   4.2\\
NGC6652 &   7.2$\pm$ 0.5 & -2.63$\pm$0.16 &  10.3 &   5.6$\pm$ 0.3 & -2.51$\pm$0.15 &   6.4\\
NGC6656 &  46.8$\pm$ 1.4 & -0.60$\pm$0.05 &  13.0 &  35.2$\pm$ 1.1 & -0.59$\pm$0.05 &  12.7\\
NGC6681 &  13.0$\pm$ 0.5 & -1.08$\pm$0.05 &  16.8 &   9.8$\pm$ 0.4 & -1.06$\pm$0.05 &  16.9\\
NGC6715 &  12.2$\pm$ 0.5 & -1.76$\pm$0.09 &  20.1 &   8.0$\pm$ 0.4 & -2.10$\pm$0.11 &  71.7\\
NGC6717 &  12.1$\pm$ 1.6 & -1.72$\pm$0.19 &  13.4 &   9.2$\pm$ 1.2 & -1.66$\pm$0.19 &  12.9\\
NGC6723 &  32.2$\pm$ 1.1 & -0.75$\pm$0.06 &   9.4 &  24.2$\pm$ 0.8 & -0.71$\pm$0.06 &   9.0\\
NGC6752 &  28.8$\pm$ 1.3 & -0.99$\pm$0.07 &  58.1 &  21.8$\pm$ 1.0 & -0.99$\pm$0.07 &  57.7\\
NGC6779 &  20.0$\pm$ 0.6 & -1.02$\pm$0.04 &  18.9 &  15.1$\pm$ 0.4 & -0.99$\pm$0.04 &  19.0\\
NGC6809 &  77.5$\pm$ 8.3 & -0.57$\pm$0.16 &   6.3 &  57.4$\pm$ 6.2 & -0.50$\pm$0.16 &   5.2\\
NGC6838 &  72.3$\pm$14.8 & -1.47$\pm$0.47 &   1.1 &  54.5$\pm$11.1 & -0.94$\pm$0.47 &   0.5\\
NGC6934 &  11.2$\pm$ 0.5 & -1.35$\pm$0.05 &  33.9 &   8.5$\pm$ 0.4 & -1.31$\pm$0.05 &  33.9\\
NGC6981 &  22.8$\pm$ 0.8 & -0.66$\pm$0.06 &  13.8 &  17.0$\pm$ 0.6 & -0.63$\pm$0.06 &  12.9\\
NGC7078 &  17.1$\pm$ 0.5 & -0.98$\pm$0.04 & 163.0 &  12.9$\pm$ 0.4 & -0.97$\pm$0.04 & 161.9\\
NGC7089 &  18.3$\pm$ 0.6 & -1.02$\pm$0.06 &  45.9 &  13.8$\pm$ 0.4 & -1.01$\pm$0.06 &  45.5\\
NGC7099 &  16.1$\pm$ 0.7 & -1.04$\pm$0.05 &  80.4 &  12.2$\pm$ 0.5 & -1.03$\pm$0.05 &  79.7\\

\hline
\end{tabular}
\label{table1}
\end{table*}
\end{center}

\bibliographystyle{apj}
\bibliography{mseg}

\end{document}